\documentclass[10pt]{article}
\usepackage[utf8]{inputenc}
\usepackage[T1]{fontenc}
\usepackage{amsmath}
\usepackage{amsfonts}
\usepackage{amssymb}
\usepackage{mhchem}
\usepackage{stmaryrd}
\usepackage{graphicx}
\usepackage[export]{adjustbox}
\usepackage{float}
\usepackage{booktabs}
\graphicspath{ {./images/} }
\usepackage{hyperref}
\hypersetup{colorlinks=true, linkcolor=blue, filecolor=magenta, urlcolor=cyan,}
\title{Cap or No Cap? What Can Governments Do to Promote EV Sales?}

\author{Zunian Luo \thanks{Weston High School, Weston, MA; Massachusetts Institute of Technology JTL Urban Mobility Laboratory; Columbia University Department of Economics.}}
\date{October 2021}

\begin{document}
\maketitle

\begin{abstract}
This paper examines the effect of the federal EV income tax subsidy on EV sales. I find that reduction of the federal subsidy caused sales to decline by $43.2 \%$. To arrive at this result, I employ historical time series data from the Department of Energy Alternative Fuels Data Center. Using the fact that the subsidy is available only for firms with fewer than 200,000 cumulative EV sales, I separate EV models into two groups. The treatment group consists of models receiving the full subsidy, and the control group consists of models receiving the reduced subsidy. This allows for a difference in differences (DiD) model structure. To examine the robustness of the results, I conduct regression analyses. Due to a relatively small sample size, the regression coefficients lack statistical significance. Above all, my results suggest that federal incentives designed to promote EV consumption are successful in their objectives.
\end{abstract}
\vspace{1em}

\newpage

\section{Introduction}
Supporting the rise in popularity of plug-in hybrid electric vehicles (PHEVs), and more recently, battery electric vehicles (BEVs), federal and state policymakers have enacted various electric vehicle (EV) purchase incentives aimed at reducing carbon emissions. However, the federal EV income tax subsidy---the United States' largest EV purchase incentive---is controversial in part because of the lack of empirical evidence.

Existing EV purchase incentives come in many types, including sales tax exemptions, income tax credits, and tax rebates. Policies may target retirement and replacement decisions (e.g., "cash for clunkers") or specific vehicle technologies, such as by offering different subsidies for BEVs and PHEVs (Gayer and Parker, 2013; Li, Linn, and Spiller, 2013; DeShazo, 2016). Smaller-scale state subsidies vary widely, from tax rebates of up to $\$ 3,500$ in Colorado to no subsidies in Arkansas (AFDC, 2021). Non-pecuniary incentives such as free parking, high occupancy vehicle lane access, and reduced toll fees have also been implemented (Ajanovic and Haas, 2016; Clinton et al., 2015).

However, the most prominent EV subsidy in the United States is the federal EV income tax credit granted by the American Clean Energy and Security Act of 2009. It provides a minimum credit of $\$ 2,500$, up to $\$ 7,500$, based on the vehicle's battery capacity and gross vehicle weight rating (American Clean Energy and Security Act, 2009). To qualify, either a PHEV or a BEV must: (1) have a battery capacity of no less than five kilowatt-hours, (2) use an external source of energy to recharge the battery (plug-in), (3) have a gross vehicle weight rating not exceeding 14,000 pounds, and (4) meet specified emissions standards (AFDC, 2021). Importantly, the tax credit is cut to $50 \%$ of the original value in the second calendar quarter following the quarter in which the automaker passed the $200,000 \mathrm{EV}$ sales threshold. It is cut down to $25 \%$ after another two quarters and finally to $0 \%$ after another two quarters. The federal subsidy has been fully eliminated for Tesla and GM, who exceeded the sales cap in $Q 3$ and Q4 of 2018, respectively (see Figure 1). The sales cap is a natural way to evaluate the effect of the federal subsidy on EV consumption as it creates an environment where some automakers do and do not qualify.

\begin{figure}[!ht]
\includegraphics[max width=\textwidth]{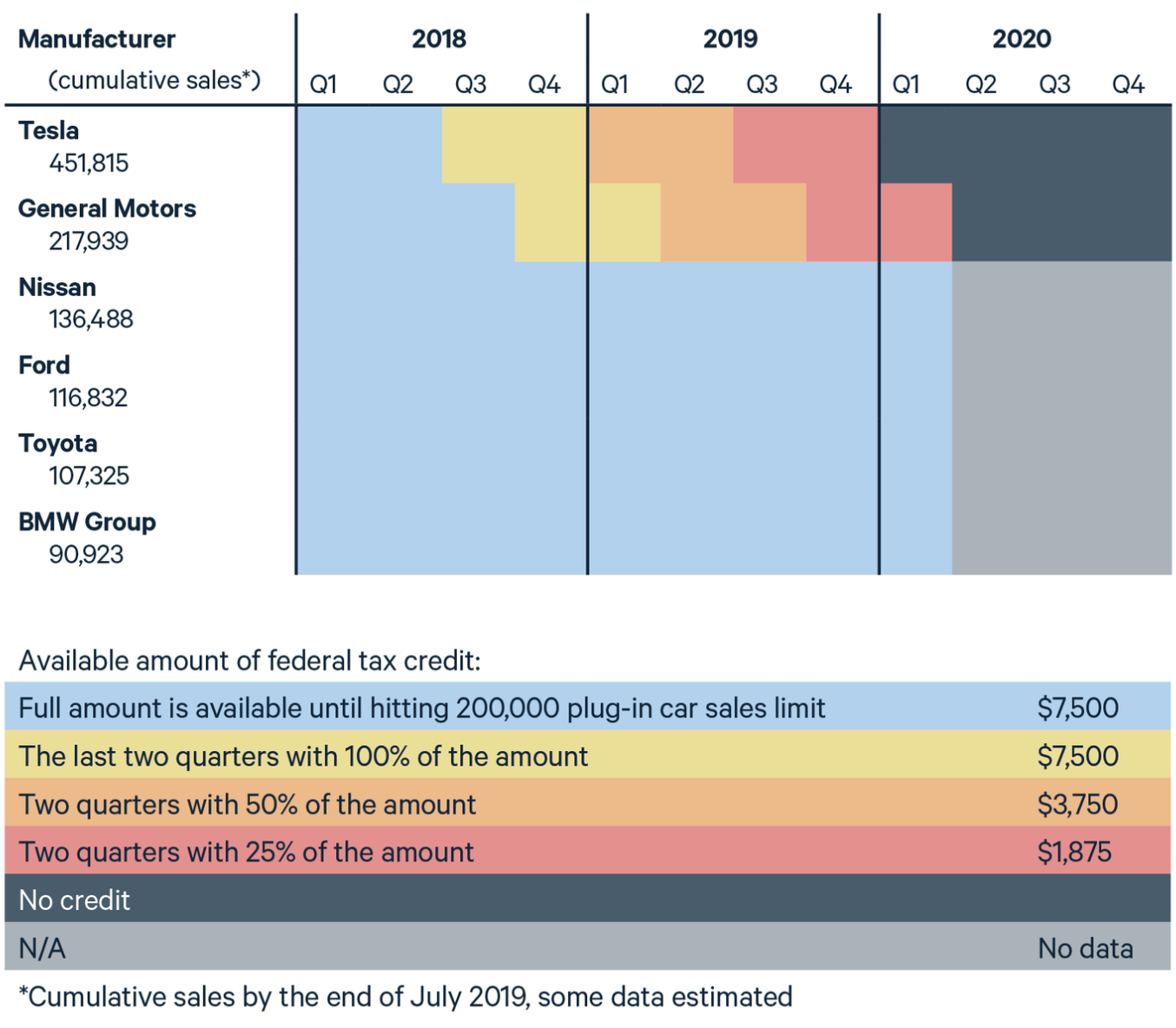}

\caption[Caption for LOF]{Phase-out Schedule of U.S. Federal EV Income Tax Credit.\footnotemark \ Credit: Resources for the Future}

\end{figure}

As the climate crisis worsens, policymakers have sought to cut emissions in the transportation sector-the fastest-growing emitter of carbon dioxide worldwide and the source of $29 \%$ of the United States' carbon emissions (Environmental Protection Agency, 2019; World Resources Institute, 2019). The 2021 Infrastructure Investment and Jobs Act recognizes the need to reduce emissions. The bill dedicates $\$ 7.5$ billion to building EV infrastructure and, if enacted, will be the "largest investment in clean energy transmission and EV infrastructure in [U.S.] history" (The White House, 2021). In addition, the Clean Energy for America Act proposes to increase the maximum credit of the current federal income tax subsidy from $\$ 7,500$ to $\$ 12,500$, while the Electric CARS Act proposes to remove the existing sales cap (Clean Energy for America Act, 2021; Electric CARS Act, 2021). Given the recent uptick in congressional interest for EV policy, an empirical analysis of the federal subsidy's effect on consumer behavior may be especially pertinent to lawmakers.

\footnotetext{The “No data” areas in Figure 1 can be ignored because Tesla and GM are the only two automakers that have sold more than 200,000 EVs at the time of writing.}

Proponents of these subsidies argue EVs generate a variety of long-term and short-term benefits, including enhanced energy security, reduced negative environmental externalities, and benefits to other EV producers in the form of innovation spillovers (Levitt, List, and Syverson, 2013; AFDC, 2021). Bloom, Schankerman, and van Reenen (2013) estimate that positive innovation spillovers cause social returns to research and development (R\&D) that exceed private returns, providing a strong case for government support in R\&D. Meanwhile, opponents of EV subsidies argue that "free-riding" from wealthy buyers compromises their cost-effectiveness, and that producers may simply raise prices in response to an extraneous price reduction (DeVries 2018). In this paper, a difference in differences (DiD) estimate is employed to evaluate the effect of the federal subsidy on EV sales (i.e., the treatment effect).\footnote{The "treatment" indicates a reduction of the federal subsidy for an EV automaker.} Intuitive economic theory indicates that when the price of a normal good rises, \textit{ceteris paribus}, its consumption falls. In the EV context, if an automaker no longer qualifies for the full federal subsidy, consumers will be forced to pay a higher out of pocket cost. Given the higher cost, it is expected that EV consumption will fall. The question is: by how much?

Two primary considerations motivate the analysis. First, EVs are underproduced and underconsumed relative to their socially optimal quantity because most market participants do not fully internalize their environmental benefits. Fuel savings - the primary cost advantage of EVs over ICEs - are an important factor influencing a consumer's decision to purchase an EV or an ICE. According to recent studies, most consumers fail to fully internalize the environmental benefits of EVs because they value them less than a reduction in purchase price (Ziegler, 2012; Daziano and Bolduc, 2013; Heffner, Kurani, and Turrentine, 2007).\footnote{It should be noted that contrary to the popular conception of EVs as "zero emissions" vehicles, they are do not truly produce zero emissions because they indirectly emit carbon through electricity generation. See, for example, President Biden Announces Steps to Drive American Leadership Forward on Clean Cars and Trucks, $2021$.} Gillingham, Houde, and Benthem (2021) find that consumers act myopically with respect to fuel costs, valuing a purchase price reduction about $2.6$ to $6.6$ times more than the equivalent amount in fuel savings. An example of the "energy efficiency gap," unobserved costs and benefits of EV ownership are another reason consumers are unable to fully internalize environmental benefits (an example of the "energy efficiency gap") (Greenstone and Allcott, 2012).\footnote{While policymakers often perceive energy efficient technologies as a win-win opportunity, the existence of unobserved costs and benefits (which are difficult for buyers to examine) leads buyers to become imperfectly informed about energy efficient investments. This information asymmetry manifests in investment inefficiency (i.e., the so-called "energy efficiency gap")} This literature suggests that direct subsidies---often more salient than intangible environmental benefits or long-term fuel savings may be necessary to increase consumption and production of EVs to their socially optimal quantity.

Second, subsidies could be needed to counteract upward price pressure caused by unfavorable supply-side factors such as stubbornly low oil prices and high lithium-ion battery costs. While there have been numerous projections of the demise of nonrenewable fossil fuels, it is unlikely that all fossil fuels will be depleted soon.\footnote{See, for example, When Fossil Fuels Run Out, What Then?, $2019$.} Reasons include the invention of new extraction technologies, the availability of large deposits of oil shales, and the increasing success rate of development and exploratory wells in recent decades. Oil shales are a telling sign that fossil fuels are here to stay; if they become commercially viable, they could more than triple the world's existing oil reserves (Dyni, 2006; Covert, Greenstone, and Knittel, 2016). Despite the falling price of lithium-ion batteries, a recent study by Bloomberg NEF (2021) suggests that battery prices must fall below $\$ 100$ per kWh (from their current price of around $\$ 140 / \mathrm{kWh}$) for manufacturers to produce EVs at a cost-competitive price with ICEs (Ziegler and Trancik, 2021). Even as the declining price of lithium-ion batteries and increasing competition in the EV market give reasons for hope, the non-trivial possibility of a substantial increase in oil reserves means that subsidies may be necessary to stimulate EV sales.

\section{Methods}
\subsection{Data Collection}
I obtain observational data from the United States Department of Energy Alternative Fuels Data Center (AFDC), officially titled "U.S. Plug-In Electric Vehicle Sales by Model."\footnote{The Alternative Fuels Data Center is a reputable government source that provides data and general information about alternative fuels and their federal and state subsidies.} The dataset contains annual EV sales by model and type (BEV/EV or PHEV) from 2011-2019 for all EV manufacturers in the United States. The sales data is reported to the Department of Energy by the Transportation Research Center at Argonne National Laboratory. I collect corporate financial statistics (net income, total assets, and total liabilities) from the balance sheets and income statements of Tesla, Ford, GM, and BMW (Standard \& Poor's Compustat, 2021).

\subsection{Difference in Differences}
As of August 2021, the difference between the average price of EVs and ICEs is $\$ 7,150$ (Kelley Blue Brook, 2021). The federal subsidy theoretically makes EVs more attractive since it reduces this gap in average purchase cost. I evaluate the extent to which the subsidy increases sales, or equivalently, the extent to which reduction of the subsidy decreases sales. The $200,000$ EV sales cap is an exogenous source of variation that can be used to assess the effect of the federal subsidy on EV sales.

This method assumes that the differences in pre-treatment sales and financial control variables between the treatment and control units remain time-invariant. The treatment effect can then be found by calculating the difference in the change in average sales before and after the effective treatment date between the treatment and control unit (see Table 1). Often called "difference-in-differences" (DiD), this method identifies the effect of the treatment (reduction in the federal subsidy) by controlling for pre-treatment differences among the observed units and their manufacturers (Card and Kreuger, 1994). \\

\begin{center}
\textbf{Table 1.} Treatment and Control Unit Sales Level.
\vspace{1em}

\begin{tabular*}{\columnwidth}{@{\extracolsep{\stretch{1}}}*{7}{r}@{}}
\toprule[1.5pt] \\
Model & 2014 & 2015 & 2016 & 2017 & 2018 & 2019 \\ \\
\midrule \\
Chevrolet Volt & 18,805 & 15,393 & 24,739 & 20,349 & 18,306 & 4,915 \\ \\
Ford Fusion Energi & 11,550 & 9,750 & 15,938 & 9,632 & 8,074 & 7,476 \\ \\
Tesla Model S & 16,750 & 26,200 & 30,200 & 26,500 & 25,745 & 15,090 \\ \\
BMW i3 & 6,092 & 11,024 & 7,625 & 6,276 & 6,117 & 4,854 \\ \\
\bottomrule[1.5pt]
\end{tabular*} \\
\end{center}

\noindent
Note: The Chevrolet Volt and Tesla Model S are treatment units, and the Ford Fusion Energi and BMW i3 are control units. Credit: AFDC. \\

The treatment companies in this study were Tesla and General Motors (GM) since they were the only automakers that exceeded the sales cap and thus no longer qualify for the full federal subsidy.\footnote{2019 is the post-treatment year in which GM and Tesla receive the federal subsidy below the original amount. See Figure $1$.} Their flagship EV units are the Tesla Model S and Chevrolet Volt (a GM model), motivating the analysis of the subsidy's effect using sales of these two units. The two corresponding control companies are Ford and BMW, and their flagship EV units are the Ford Fusion Energi and BMW i3. Given similar characteristics such as price and type (e.g., BEV, PHEV), I demonstrate that the BMW i3 BEV is a valid control for the Tesla Model S BEV, and the Ford Fusion Energi PHEV is a valid control for the Chevrolet Volt PHEV.

To measure the treatment effect using this model, several assumptions are made: (1) the anticipation of the treatment among market participants has a negligible effect on sales, (2) there are no spillover effects, (3) manufacturer financial statistics (net income, total assets, and total liabilities) exhibit parallel trends, and (4) sales of the treatment and control units follow parallel trends.

\subsection{Econometric Model}

\subsubsection{Difference in Differences Estimate}

A difference in differences estimate is conducted using control units and treatment units that are similar in price and type, and whose automakers have similar financial statistics. The subsidy shifts sales by $\widehat{\delta_{1}}$, as defined below:

\begin{equation}
\widehat{\delta_{1}}=\left(\bar{y}_{B, 2}-\bar{y}_{B, 1}\right)-\left(\bar{y}_{A, 2}-\bar{y}_{A, 1}\right)
\end{equation}

Where car units are indexed by treatment status $\mathrm{T}=\mathrm{A}, \mathrm{B}$. A specifies the unit that is not affected by the treatment, i.e., the control unit, and B indicates the unit that is affected by the treatment, i.e., the treatment unit. Treatment and control units are observed in two time periods, $\mathrm{t}=1,2$, where 1 represents the period before the treatment goes into effect, i.e., pre-treatment, and 2 defines the period after the treatment goes into effect, i.e., posttreatment. $\bar{y}_{A, 1}$ and $\bar{y}_{A, 2}$ represent the average annual sales of the control unit before and after the effective treatment date, respectively, and $\bar{y}_{B, 1}$ and $\bar{y}_{B, 2}$ represent the corresponding average annual sales for the treatment unit. To find the sales growth of unit $\mathrm{A}$ and unit $\mathrm{B}$, the difference between unit A's average pre and post-treatment sales $\left(\bar{y}_{A, 2}-\bar{y}_{A, 1}\right)$ and the difference between unit B's average pre and post-treatment sales $\left(\bar{y}_{B, 2}-\bar{y}_{B, 1}\right)$ are calculated. Then, the treatment effect $\left(\widehat{\delta_{1}}\right)$, is found by taking the difference between the change in the treatment unit's average sales and the change in the control unit's average sales.

\subsubsection{Generalized Difference in Differences Regression}
I conduct agnostic DiD regression analyses to evaluate the robustness of the initial estimate. While the initial estimate uses data from only two pairs of treatment and control units, the regression analyses draw data from all units in the U.S EV market.\footnote{Recall that the treatment and control unit pairs used in the initial DiD estimate are (1) Tesla Model S and BMW i3, and (2) Chevrolet Volt and Ford Fusion Energi.} The sales outcome $\left(Y_{m f t}\right)$ is defined by the following linear regression model:

\begin{equation}
Y_{m f t}=\beta_{0}+\gamma_{f}+\lambda_{t}+\beta_{1} D_{f t}+\epsilon_{m f t}
\end{equation}

Where the sales outcome, $Y_{m f t}$, is stated in terms of a particular model (m) of a particular firm (f) in a particular year (t). $D_{f t}$ is a dummy variable indicating treatment status; $D_{f t}$ is 1 if a particular firm is receiving $100 \%$ of the federal subsidy in a given year and 0 if it is not.\footnote{$D_{f t}=0$ only during the post treatment period of a treatment unit. In the context of this study, $D_{f t}=0$ in the posttreatment period of Tesla and GM units and $D_{f t}=1$ in all other circumstances.} $\epsilon_{m f t}$ represents idiosyncratic noise - variables other than the federal subsidy that may affect sales of the treatment or control unit in the post treatment period. The coefficients of each firm correspond to firm fixed effects $\left(\gamma_{f}\right)$, and coefficients of each year correspond to year fixed effects $\left(\lambda_{t}\right) \cdot \beta_{1}$ indicates the treatment effect, and $\beta_{0}$ is a constant. $\beta_{1}$ and $\widehat{\delta_{1}}$ (see Equation 1) have identical interpretations. Notably, though, they can have different values because the regression model utilizes sales data from all units in the AFDC dataset, while the initial estimate includes sales data from only two pairs of treatment and control units (over a shorter period of time).\footnote{The observed time period in the initial estimate begins in 2015 because some models do not have recorded sales data prior to 2014 for some models (see Figure 2 A-B).} The assumptions (see Methods) are accounted for in Equations 2 and 3 through the assumption that the error term $\left(\epsilon_{m f t}\right)$ is uncorrelated with respect to $D_{f t}$.

\subsubsection{Logarithmic-Linear Difference in Differences Regression}
A major limitation of the linear regression is that the right skewed nominal sales data may compromise the accuracy of the regression coefficients (see Appendix C). To account for this potential shortcoming, I transform sales from nominal to logarithmic form and conduct a logarithmic-linear regression using logarithmic sales:

\begin{equation}
\log Y_{m f t}=\beta_{0}+\gamma_{f}+\lambda_{t}+\beta_{1} D_{f t}+\epsilon_{m f t}
\end{equation}

Where $\log Y_{m f t}$ is an approximation of the percentage change in sales. Equation 2 and Equation 3 are equivalent apart from the difference in functional form-that is, whether the sales outcome $\left(Y_{m f t}\right)$ is expressed in logarithmic or nominal terms.

The accuracy of the predicted treatment effect can be increased if the $\beta_{1}$ coefficient is converted from log points to percentage points. Under the central assumption that the error term, $\epsilon_{m f t}$, is not correlated with $D_{f t}$, the exact percentage change in $Y_{m f t}$ for $\Delta D_{f t}$ is evaluated:
\begin{equation}
\% Y_{m f t}=\left(\exp \left(\Delta D_{f t} \beta_{1}\right)-1\right) \times 100
\end{equation}

Where $\Delta D_{f t}$ is the change in the dummy variable $(=1-0)$ and $\beta_{1}$ is the treatment effect (expressed in log points). The percentage change in $Y_{m f t}$ approximated by $\beta_{1}$ is always a lower bound of the actual percentage change in $Y_{m f t} \cdot$\footnote{For example, a $\beta_{1}$ value of $0.50$ in the log linear regression means that a one-unit change in $D_{f t}$ is responsible for a $50 \%$ increase in $Y_{m f t}$. Converting 50 log points to percentage points via Equation 4 indicates the exact percent change in $Y_{m f t}$ is $65 \%(\approx(\exp (0.50)-1) \times 100)$.} The discrepancy between the approximation and exact percentage change in $Y_{m f t}$ associated with $\Delta D_{f t}$ rises exponentially as the $\beta_{1}$ coefficient increases.

\subsubsection{Regression Hypotheses}

The hypotheses in the linear and log-linear regression models are defined below:\\

$H_{0}: \beta_{1}=0$\\

$H_{1}: \beta_{1} \neq 0$\\

Where the significance level is $5 \% \ (\alpha=0.05)$. $H_{0}$ states that the federal subsidy has no effect on EV sales, and $H_{1}$ states that the federal subsidy has a non-zero effect on EV sales.

\section{Results}
\subsection{Parallel Sales Trends}

Several assumptions are evaluated, with the most integral being the parallel sales trends condition. Importantly, this condition can never be fully validated because the treatment unit counterfactual-the sales of the treatment unit in the post treatment period if it still had been receiving the full federal subsidy-is never observed. Instead, the condition is met if the observed outcome of the treatment and control unit visibly follow parallel trends in the pre-treatment period.

As Figure 2A shows, the sales trends of the Chevrolet Volt and Ford Fusion Energi match closely throughout the entire pre-treatment period (2015-2018), exemplifying parallel trends. Conversely, Figure 2B shows that the sales growth of the Tesla Model S and BMW i3 diverge from 2016-2017. The parallel trends condition for this pair is reasonably met since the Model S and i3 experience nearly identical sales growth from 2017-2018.\\

\begin{figure}[H]

\includegraphics[max width=\textwidth]{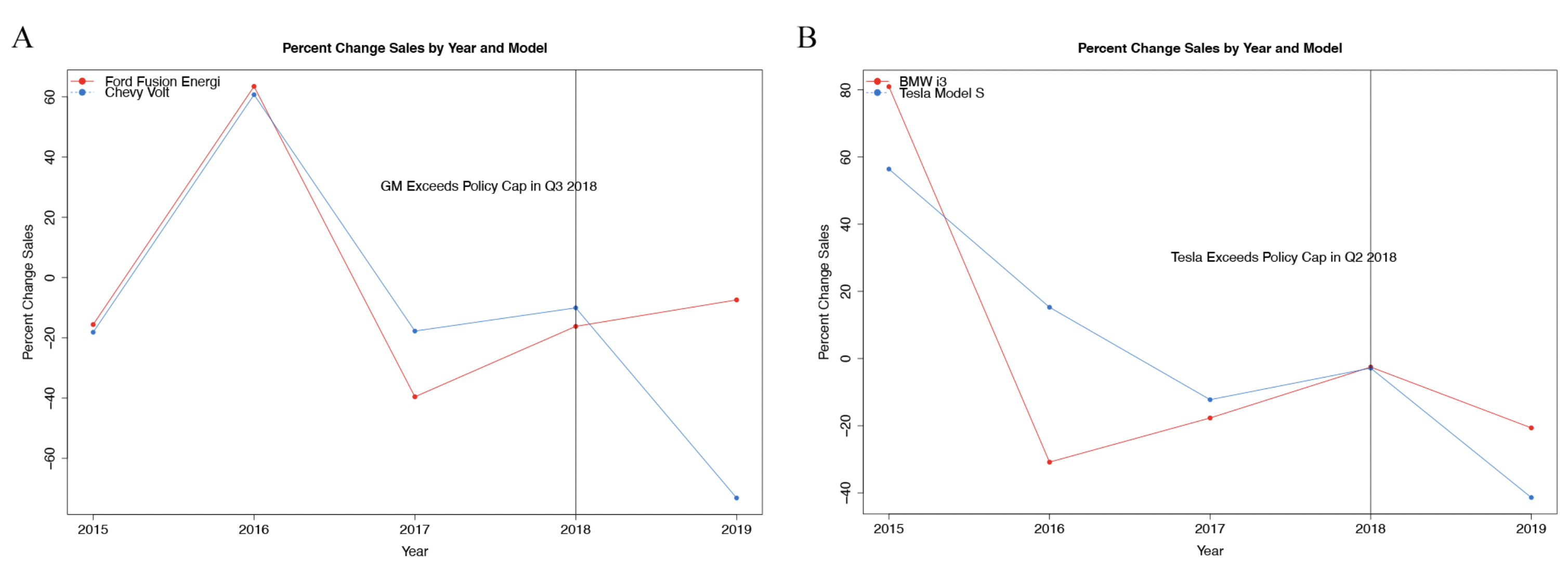}

\caption{Yearly Percentage Change in Sales. Note: A compares the sales trends of the Chevy Volt and Ford Fusion Energi, and B compares the sales trends of the Tesla Model S and BMW i3.}

\end{figure}

\subsection{Parallel Trends in Corporate Financial Statistics}

Figure 3 (A-F) demonstrates that the annual percentage change in three financial control variables approximately follow parallel trends. As seen in Figure 3E, Ford's 2,821\% decrease in net income from 2019-2020 is a clear outlier. It can be ignored because only trends in the pre-treatment period (2014-2018) are relevant to the parallel trends condition. BMW, GM, and Ford generally experienced moderate growth (often between $0-15 \%$) with respect to the three observed financial statistics throughout the entire pre-treatment period. This pattern is perhaps explained by their legacy status in the automotive market. On the other hand, Tesla's outlier trend might suggest that the chosen control firm (BMW) is invalid. The existence of Tesla as an outlier is outweighed, however, by robust parallel trends in sales of the treatment and control units and parallel trends in the financial control variables of the three other automakers.

\begin{figure} [H]

\includegraphics[max width=\textwidth]{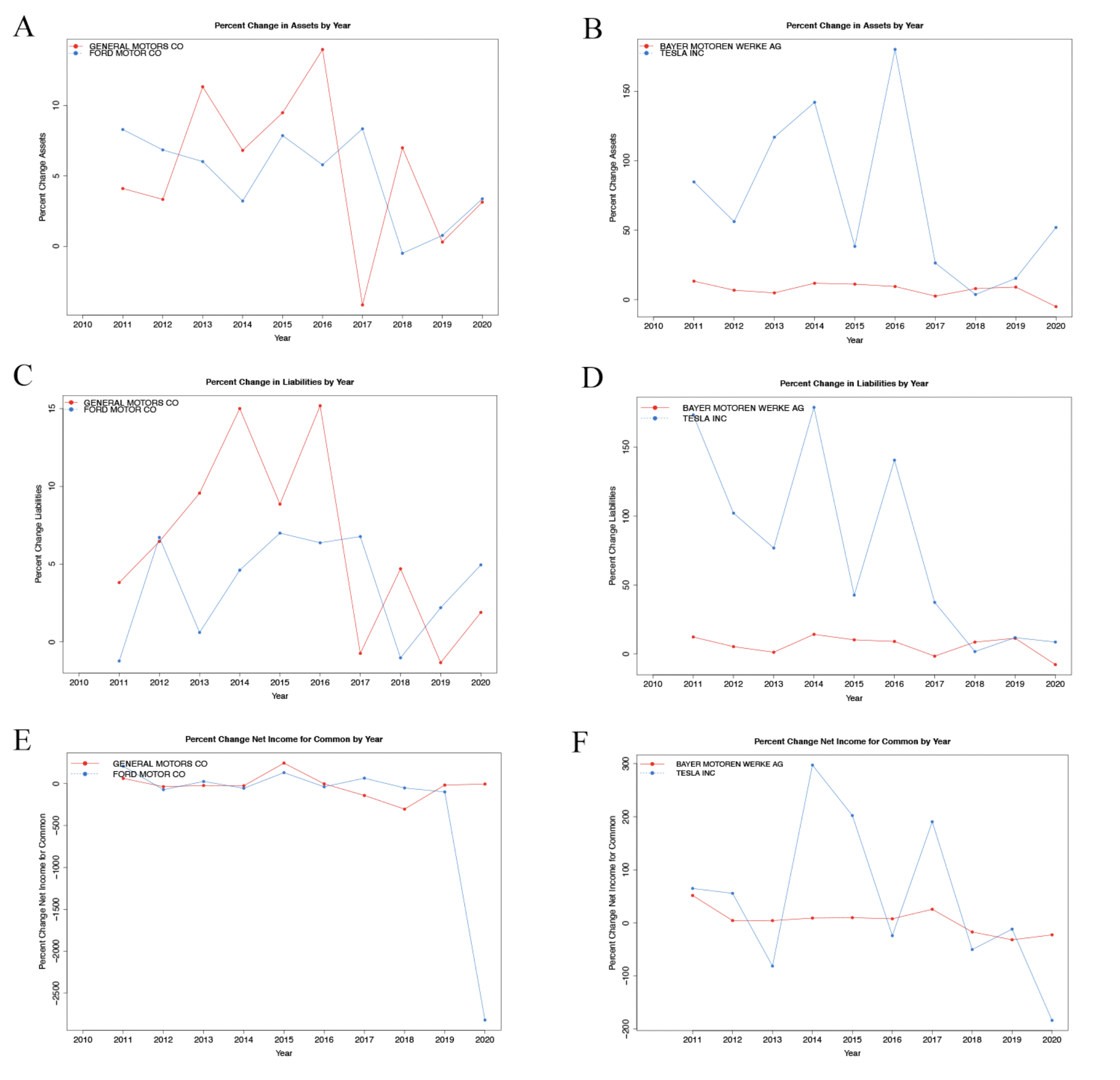}

\caption{Yearly Percentage Changes in Corporate Financial Statistics. Note: A and B show total asset trends, C and D show the total liability trends, and E and F show net income trends. A, C, and E compare GM and Ford, while B, D, and F compare Tesla and BMW.}

\end{figure}

\subsection{Effect of Federal Subsidy on Sales of Treatment Units}

\subsubsection{Initial Difference in Differences Estimate}

With the parallel trends condition largely met with respect to sales and, to a lesser degree, corporate financial statistics, the initial DiD estimate is calculated:
\begin{equation}
\widehat{\delta_{1}}=\left(\bar{y}_{B, 2}-\bar{y}_{B, 1}\right)-\left(\bar{y}_{A, 2}-\bar{y}_{A, 1}\right)
\end{equation}
As seen in the second to last column in Table 2, the reduction in the federal subsidy caused nominal sales of the Tesla Model S to decline by 11,091 and of the Chevrolet Volt to decline by 7,416, in 2019. The nominal sales decline of the Model S is $49.6 \%$ greater than that of the Chevrolet Volt. The last column shows that the nominal effect of the subsidy $\left(\widehat{\delta_{1}}\right)$ is $56.8 \%$ and $29.6 \%$ of average sales in the pre-treatment period for the Model S and Volt, respectively. Notably, the relative treatment effect $\left(\left|\widehat{\delta}_{1}\right| / \bar{y}_{B, 1}\right)$ for the Model S is $91.9 \%$ greater than for the Chevrolet Volt. The average nominal effect of the subsidy $\left(\widehat{\delta_{1}}\right)$ is 9,254, and the average relative effect is $43.2 \% \ (=$ $56.8 \%$ and $29.6 \%)$.\footnote{The relative effect expresses the treatment unit's decline in average post treatment sales as a share of average pretreatment sales.} \\

\begin{center}
\textbf{Table 2.} Difference in Difference Estimate Components.
\vspace{1em}

\begin{tabular*}{\columnwidth}{@{\extracolsep{\stretch{1}}}*{7}{r}@{}}
\toprule[1.5pt] \\
\quad & $\bar{y}_{A, 2}$ & $\bar{y}_{A, 1}$ & $\bar{y}_{B, 2}$ & $\bar{y}_{B, 1}$ & $\widehat{\delta}_{1}$ & $|\widehat{\delta}_{1}| / \bar{y}_{B, 1}$ \\ \\
\hline \\
Chevrolet Volt \& Ford Fusion Energi & 7,476 & 10,989 & 4,915 & 19,518 & -11,091 & 0.568 \\ \\
Tesla Model S \& BMW i3 & 4,854 & 7,427 & 15,090 & 25,079 & -7,416 & 0.296 \\ \\
\bottomrule[1.5pt]
\end{tabular*} \\
\end{center}

\noindent
Note: $\bar{y}_{A, 1}$ and $\bar{y}_{A, 2}$ represent the average annual sales for the control units (Ford Fusion Energi and BMW i3) before and after the treatment went into effect, respectively. $\bar{y}_{B, 1}$ and $\bar{y}_{B, 2}$ represent the corresponding average annual sales for the treatment units (Chevrolet Volt and Tesla Model S). $\widehat{\delta_{1}}$ is the treatment unit sales lost due to the reduction of the federal subsidy, and $\left|\widehat{\delta}_{1}\right| / \bar{y}_{B, 1}$ represents $\widehat{\delta}_{1}$ as a percentage of the treatment unit's average pre-treatment sales.

\subsubsection{Generalized Difference in Differences Regression}
Regression analyses with more rigorous statistical techniques are performed in order to determine the robustness of the predicted treatment effect. The linear regression model is specified by Equation 2:
\begin{equation}
Y_{m f t}=\beta_{0}+\gamma_{f}+\lambda_{t}+\beta_{1} D_{f t}+\epsilon_{m f t}
\end{equation}
As seen in Table 3, the regression 1 $\beta_{1}$ coefficient of $-16575.2$ means the subsidy caused sales to decline by about 16572 units. Among the specifications, regression 1 has the greatest absolute t-value (4.493) and is the most statistically significant $(\mathrm{p}$-value $=9.05 \mathrm{e}-06)$. The range between the smallest and largest $\beta_{1}$ coefficient in absolute terms is $14740.9(=16,575.2-1,834.3)$. The range between the largest and smallest t-value in absolute terms is $3.347\left(=4.493-1.146\right.$ ). Regressions $1-5$ are statistically significant ( $\mathrm{p}$-value $<0.05$ ). The absolute value of the $\beta_{1}$ coefficient decreases as the number of years in the regression decreases. Meanwhile, the p-value of the $\beta_{1}$ coefficient increases as the number of years in the regression decreases (except for regression 7). \\

\begin{center}
\textbf{Table 3.} Linear Regression Coefficients.
\vspace{1em}

\begin{tabular*}{\columnwidth}{@{\extracolsep{\stretch{1}}}*{7}{r}@{}}
\toprule[1.5pt] \\
Independent Variable & Estimate $\left(\beta_{1}\right)$ & Standard Error & $\mathrm{t}-\mathrm{value}$ & $\operatorname{Pr}(>|\mathrm{t}|)$ \\ \\
\hline \\
Subsidy Eligible (Regression 1) & -16,575.2 & 3,689.5 & -4.493 & 9.05e-06 \\ \\
Subsidy Eligible (Regression 2) & -15,641.2 & 3,885.0 & -4.026 & 6.86e-05 \\ \\
Subsidy Eligible (Regression 3) & -14,713.8 & 4,122.3 & -3.569 & 0.000412 \\ \\
Subsidy Eligible (Regression 4) & -13,731.71 & 4,424.12 & -3.104 & 0.00211 \\ \\
Subsidy Eligible (Regression 5) & -11,958.0 & 4,773.7 & -2.505 & 0.01299 \\ \\
Subsidy Eligible (Regression 6) & -9,399.2 & 5,182.9 & -1.813 & 0.0716 \\ \\
Subsidy Eligible (Regression 7) & -6,227.2 & 5,435.0 & -1.146 & 0.2544 \\ \\
Subsidy Eligible (Regression 8) & 1,834.3 & 1,401.3 & 1.309 & 0.196172 \\ \\
\bottomrule[1.5pt] \\
\end{tabular*} \\
\end{center}

\noindent
Note: Linear regressions are conducted across 8 different year specifications. Regression 1 contains sales data for years 2011-2019, regression 2 contains sales data for years 2012-2019, regression 3 for years 2013-2019, and so on until regression 8, which includes sales data for years 2018-2019. The values in the "Estimate" column represents the $\beta_{1}$ coefficients (expressed in nominal terms). The standard deviation of $\beta_{1}$ coefficients across all specifications is $4,790.7075$ units (figure not shown in table).

\subsubsection{Logarithmic-Linear Difference in Differences Regression}
The log-linear regression is defined by Equation 3:
\begin{equation}
\log Y_{m f t}=\beta_{0}+\gamma_{f}+\lambda_{t}+\beta_{1} D_{f t}+\epsilon_{m f t}
\end{equation}
As seen in the "Estimate" column in Table 4, the $\beta_{1}$ coefficient in regression 6 is $1.56285$. The corresponding $\% Y_{m f t}$ (377.2403) indicates the federal subsidy increases sales by about $377 \%$. The $\% Y_{m f t}$ in regression 1 $(170.3805)$ and in regression 8 (177.4609) are relatively more modest. There are no statistically significant regression coefficients among the year specifications. The range between the smallest and largest $\% Y_{m f t}$ is $206.8598 (=377.2403-170.3805)$, and the range between the smallest and largest t-value is $0.17993 (=1.393-$ 0.894). In contrast to the linear regression, the $\beta_{1}$ coefficient increases as the number of years included in the specification decreases (except for regressions 7 and 8). \\

\begin{center}
\textbf{Table 4.} Logarithmic-Linear Regression Coefficients. \\
\vspace{1em}

\begin{tabular*}{\columnwidth}{@{\extracolsep{\stretch{1}}}*{7}{r}@{}}
\toprule[1.5pt] \\
Independent Variable & Estimate($\beta _1$) & $\% Y_{mft}$ & Standard Error & t-value & Pr($>|t|$) \\ \\
\hline \\
Subsidy Eligible (Regression 1) & 0.99466 & 170.3805 & 1.11248 & 0.894 & 0.371772 \\ \\
Subsidy Eligible (Regression 2) & 1.13398 & 210.8002 & 1.15282 & 0.984 & 0.325912 \\ \\
Subsidy Eligible (Regression 3) & 1.28049 & 259.8403 & 1.17431 & 1.090 & 0.276340 \\ \\
Subsidy Eligible (Regression 4) & 1.34049 & 282.0915 & 1.17633 & 1.140 & 0.25549 \\ \\
Subsidy Eligible (Regression 5) & 1.48502 & 341.5054 & 1.15260 & 1.288 & 0.198989 \\ \\
Subsidy Eligible (Regression 6) & 1.56285 & 377.2403 & 1.1218 & 1.393 & 0.16549 \\ \\
Subsidy Eligible (Regression 7) & 1.35010 & 285.7811 & 1.0020 & 1.347 & 0.180708 \\ \\
Subsidy Eligible (Regression 8) & 1.02051 & 177.4609 & 0.9964 & 1.024 & 0.310397 \\ \\
\bottomrule[1.5pt] \\
\end{tabular*} \\
\end{center}

\noindent
Note: Log-linear regressions are conducted across 8 different year specifications. Regression 1 contains sales data for years 2011-2019, regression 2 contains sales data for years 2012-2019, regression 3 for years 2013-2019, and so on until regression 8, which includes sales data for years 2018-2019. The "Estimate" column includes the unmodified $\beta_{1}$ coefficients. The values in the " $\% Y_{m f t} "$ column are the outputs from Equation 4. The standard deviation of $\% Y_{m f t}$ across all specifications is $80.6274 \%$ (figure not shown in table).

\section{Discussion}

\subsubsection{Difference in Differences Estimate}

Losing the federal subsidy caused an average sales decline equivalent to $43.2 \%$ of average pre-treatment sales (Table 2); the federal subsidy has a substantial positive impact on EV sales.\footnote{Recall that $\widehat{\delta_{1}}$ is $56.8 \%$ for the Chevrolet Volt and $29.6 \%$ for the Tesla Model S.} Notably, the relative effect of the subsidy on sales of the Volt is $91.9 \%$ greater compared to the Model S. This result is consistent with Layard, Mayraz, and Nickell (2008), who found that lower-income consumers have a higher marginal utility of income than higher-income consumers. Since the more affordable Volt targets middle-income consumers, while the luxury Model S is geared towards higher-income consumers, the same price reduction should engender a stronger motivation to purchase the former than the latter.

Moreover, there is reason to believe the true treatment effect is greater than the initial estimate suggests. While a reduced form of the subsidy was available for GM and Tesla throughout 2019, the subsidy has since been fully eliminated for the two companies. For GM in particular, 2019 is an incomplete post-treatment period as the firm was still receiving $100 \%$ of the subsidy in Q1 of 2019.\footnote{The issue of an incomplete post treatment period could have been at least partly avoided if sales data were organized by month or quarter. However, AFDC dataset used in this study organizes sales by year.} This likely caused GM's EV sales in the post-treatment period to be slightly greater than if the treatment was in effect for the entire post-treatment period (i.e., if GM was following the same phase out schedule as Tesla). Unlike GM, Tesla was receiving the reduced subsidy throughout the post-treatment period; the treatment effect may be underestimated to a greater degree for the Volt relative to the Model S (see Figure 1).

\subsubsection{Generalized Differences in Differences Regression}
The fact that many of the linear regression specifications have negative $\beta_{1}$ coefficients runs against the grain of intuitive economic theory.\footnote{It is a well-known among experts and laymen that demand for a normal good increases as its price decreases (as opposed to say, Veblen or Giffen goods). However, Table 3 shows that the federal subsidy decreases sales.} This factor, in addition to the right skewed nominal sales data, reduces confidence in the results of the linear regression and motivates a logarithmic linear regression analysis (see Appendix C).\footnote{The counterfactual's significant sensitivity to the number of years in the pre-treatment period reduces confidence in the $\beta_{1}$ coefficients of both the linear and log-linear regression.} The substantial positive treatment effect indicated by the log-linear regression is in line with that of the initial estimate. The log-linear regression shows that the federal subsidy increases EV sales by more than $300 \%$ in the most extreme case and by about $170 \%$ in the least extreme case. It is highly implausible that the federal subsidy increases EV sales to the extent that the log-linear regression indicates.

The $\beta_{1}$ coefficients of linear regression specifications $1-5$ are statistically significant at the $5 \%$ level, favoring $H_{1}$. T high standard errors across all log-linear specifications do not support $H_{1}$. However, they suggest that more modest $\beta_{1}$ coefficients could be obtained if additional data points are included in the regression. The overall lack of statistical significance in the results of the regression model give reason for greater reliance on the results of the initial estimate.

\subsection{Limitations}
\subsubsection{Availability of Data}
The lack of statistical significance in the regression model is not so much a result of underlying flaws in study design as lack of sales data (especially with respect to treatment units). Several contingent factors motivate future empirical evaluations of the treatment effect. Variation in $\beta_{1}$ will decrease if data from more control and treatment units is utilized. While a growing $\mathrm{EV}$ market clearly attracts new automakers to enter (creating additional control units), it also motivates some existing manufacturers to produce new car models even if they do not qualify for the full federal subsidy (creating additional treatment units). Additionally, several EV automakers are expected to exceed the sales cap in the next few years (Federal EV Tax Credit Phase Out Tracker by Automaker, 2021).\footnote{Toyota is expected to exceed 200,000 EV sales in Q2 of 2022, followed by Ford in Q3 of 2022.} If they receive less than the full federal subsidy, their existing (control) units would become treatment units. Variation in $\beta_{1}$ can also be reduced by utilizing more frequent sales data (e.g., monthly, weekly).\footnote{The latest year of available sales data in the AFDC dataset is 2019, meaning each treatment and control unit has only one data point from the post-treatment period. If, for example, the AFDC records sales data on a monthly basis, the empirical analyses specified in this study can draw from a data pool of post-treatment units 12 times larger than the size of the data pool at the time of writing.} Even if the AFDC dataset continues to be organized at the annual level, the number of data points will naturally increase as data is added from more years. It would be insightful for further empirical examinations of the federal subsidy's effect to incorporate more frequent sales data from more units.

\subsubsection{Study Scope}
It is possible the treatment effect found in the initial estimate $\left(\widehat{\delta_{1}}\right)$ differs significantly from the average treatment effect for all EV manufacturers. If this is the case, the application of the treatment effect would be limited to Tesla and GM. However, given that data for treatment units is limited to Tesla and GM, the fact that the treatment effect $\left(\widehat{\delta_{1}}\right)$ is broadly consistent with intuitive economic theory suggests that it should apply to all EV manufacturers.

The results of this study are only applicable to the U.S federal income tax credit. One may intuit that a proportional relationship exists between (1) the effect of smaller state income tax credits and the larger U.S federal income tax credit, and (2) between other countries' federal income tax credits and the U.S federal income tax credit. But, the question of whether such proportional relationships exist unclear. Future research that seeks to answer these questions would provide insight into the relationship between federal and state EV subsidy design.\footnote{The question of whether relationship (2) is proportional is especially important given the expected rise in India and China's share of global energy consumption, of which transportation is a major driving force. See China and India account for half of global energy growth through 2035, 2011.}

\subsection{Omitted Variables}
\subsubsection{Assumptions}

The potential for factors unrelated to the federal subsidy to have affected EV sales in the post-treatment period presents a possible limitation of this study. For instance, there are at least three potential limitations concerning the previously stated assumptions (see Methods). First, the loss of the subsidy may have caused a subgroup of price-conscious consumers to strategically shift away from the treatment unit in favor of the control unit, violating assumption 2.\footnote{Recall that the treatment and control units are selected based on their similarity with respect to a set of characteristics. See Methods.} While such behavior would exaggerate the difference in post-treatment sales between the treatment and control unit, the potential overestimation appears to be trivial. For, such strategically acting consumers are faced with a wide variety of similar EV brands. Thus, sales of the treatment unit would be diverted to a broader group of models.\footnote{Prominent alternatives to Tesla include luxury brands Mercedes, Porsche, and Volvo. Similar mid-price alternatives to Chevrolet include Nissan, Toyota, and Smart.} Second, the parallel trends condition may be violated if one of the observed units has a battery fire incident after the federal subsidy is reduced for the treatment firm. However, these incidents are not particularly relevant to the parallel trends condition. Reports of EV battery fires are not exclusive to a specific brand and the resulting consumer shock would likely affect sales of all EV models rather than a single automaker (see, for example, Lambert, 2019; Colias and Foldy, 2021). Third, the parallel trends condition may also be compromised if the error term, $\epsilon_{m f t}$, is correlated with $D_{f t}$. Nonetheless, it is appropriate to assume $\epsilon_{m f t}$ is not correlated with $D_{f t}$ since the treatment and control units are reasonably similar (see Results).\footnote{A notable feature of the treatment unit counterfactual is that it is never observed, meaning the parallel trends condition cannot be completely verified.} While the assumptions in this study are limited in some regards, closer examination reveals that they are unlikely to cause a major overestimation of the treatment effect.

\subsubsection{Anticipation of Treatment}
The treatment effect may be overstated if market participants act strategically in anticipation of the treatment. For example, producers expecting to lose the federal subsidy may decide against starting the development of new EV models, stop ongoing development of EV models, or discontinue existing EV models that are most adversely affected by a reduction in the federal subsidy. Consumers can act strategically by purchasing the treatment unit earlier than they anticipated in order to take advantage of the federal subsidy. These behaviors could be seen in a pronounced swing in sales of the treatment unit shortly before the treatment goes into effect. Figure 2 (A-B) reveals that instead of being a major swing, sales growth from 2017-2018 was the smallest for all units (excluding the Ford Fusion Energi).

\subsubsection{Discontinued Chevrolet Volt}
The initial estimate of the treatment effect on sales of the Chevrolet Volt could be an overestimation since it does not account for the fact that the Chevrolet Volt was discontinued in March 2019 (Lambert, 2018). The lack of a 2020 model may have dissuaded consumers intent on purchasing the new model. However, the recorded sale of 68 new Volt units in 2020 implies the Volt was still available for purchase throughout 2019 and 2020 (Cain, 2021). The unit's discontinuance is unlikely to substantially confound the effect of the federal subsidy.

\section{Conclusion}
While the issue of EV policy has come to the fore, there is a lack of evidence on the effect of the federal EV income tax credit on sales. Despite potential shortcomings, this study contributes an important finding to the growing federal interest in EV policy and introduces fruitful avenues for further inquiry. Most important to this study's examination, I find that a partial reduction in the federal subsidy caused EV sales to decline by $43.2 \%$ on average. This indicates the federal subsidy has a substantial positive effect on EV sales. It further suggests that one way to stimulate sales could be to raise the sales cap threshold. To build upon the results of this study, future work should use rigorous empirical techniques that incorporate more frequent time series data and data from more units.

\section{Acknowledgements}
I am very grateful to Bhargav Gopal, Mary Liu, and Sara Ashbaugh for their helpful comments, insightful discussion, and generous support. I would also like to thank Richard Li for copy editing support. 

\section{References}
\begin{sloppypar}
Aguirre, K., Eisenhardt, L., Lim, C., Nelson, B., Norring, A., Slowik, P., \& Tu, N. (2012). Lifecycle analysis comparison of a battery electric vehicle and a conventional gasoline vehicle. California Air Resource Board. \href{https://www.ioes.ucla.edu/wp-content/uploads/batteryelectricvehiclelca2012.pdf}{https://www.ioes.ucla.edu/wp-content/uploads/batteryelectricvehiclelca2012.pdf}.
\\ \\
Ajanovic, A., \& Haas, R. (2016). Dissemination of electric vehicles in urban areas: Major factors for success. \textit{Energy, 115,} 1451-1458. \href{https://doi.org/10.1016/j.energy.2016.05.040}{https://doi.org/10.1016/j.energy.2016.05.040}.
\\ \\
Allcott, Hunt, and Michael Greenstone. 2012. "Is There an Energy Efficiency Gap?" \textit{Journal of Economic Perspectives}, 26(1) : 3-28. \href{https://doi.org/10.1257/jep.26.1.3}{https://doi.org/10.1257/jep.26.1.3}.
\\ \\
Alternative Fuels Data Center - Department of Energy. (2020). Electric Vehicle Registrations by State [Excel Spreadsheet]. Retrieved from \href{https://afdc.energy.gov/data/10962.}{https://afdc.energy.gov/data/10962.}
\\ \\
Alternative Fuels Data Center - Department of Energy. (2020). U.S. Plug-In Electric Vehicle Sales by Model [Excel Spreadsheet]. Retrieved from \href{https://afdc.energy.gov/data/10567.}{https://afdc.energy.gov/data/10567.}
\\ \\
Amy Finkelstein, \textit{E-ztax}: Tax Salience and Tax Rates, \textit{The Quarterly Journal of Economics}, Volume 124, Issue 3, August 2009, Pages 969-1010, \href{https://doi.org/10.1162/qjec.2009.124.3.969}{https://doi.org/10.1162/qjec.2009.124.3.969}.
\\ \\
Anthoff, David and Emmerling, Johannes, Inequality and the Social Cost of Carbon (August 26, 2016). FEEM Working Paper No. 54.2016, Available at SSRN: \href{https://ssrn.com/abstract=2830457}{https://ssrn.com/abstract=2830457} or \href{http://dx.doi.org/10.2139/ssrn.2830457}{http://dx.doi.org/10.2139/ssrn.2830457}.
\\ \\
Bhat, C. R., Sen, S., \& Eluru, N. (2009). The impact of demographics, built environment attributes, vehicle characteristics, and gasoline prices on household vehicle holdings and use. \textit{Transportation Research Part B: Methodological, 43}(1), 1-18. \href{https://doi.org/10.1016/j.trb.2008.06.009}{https://doi.org/10.1016/j.trb.2008.06.009}.
\\ \\
Bloom, N., Schankerman, M., \& Van Reenen, J. (2013). Identifying technology spillovers and product market rivalry. \textit{Econometrica, 81}(4), 1347-1393. \href{https://doi.org/10.3982/ECTA9466}{https://doi.org/10.3982/ECTA9466}.
\\ \\
Bloomberg NEF. (2021). Electric Vehicle Outlook 2021. \href{https://about.newenergyfinance.com/electric-vehicle-outlook/}{https://about.newenergyfinance.com/electric-vehicle-outlook/}.
\\ \\
Cain, T. (2021). Chevrolet Volt Sales Figures. Retrieved 5 September 2021, from \href{https://www.goodcarbadcar.net/chevrolet-volt-sales-figures/}{https://www.goodcarbadcar.net/chevrolet-volt-sales-figures/}.
\\ \\
Card, D., Katz, L. F., \& Krueger, A. B. (1994). Comment on David Neumark and William Wascher,"Employment effects of minimum and subminimum wages: Panel data on state minimum wage laws". \textit{ILR Review, 47}(3), 487-497. \href{https://doi.org/10.1177/001979399404700308}{https://doi.org/10.1177/001979399404700308}.
\\ \\
Carleton, Tamma and Greenstone, Michael, Updating the United States Government's Social Cost of Carbon (January 14, 2021). University of Chicago, Becker Friedman Institute for Economics Working Paper No. 2021-04, Available at SSRN: \href{https://ssrn.com/abstract=3764255}{https://ssrn.com/abstract=3764255} or \href{http://dx.doi.org/10.2139/ssrn.3764255}{http://dx.doi.org/10.2139/ssrn.3764255}.
\\ \\
Chetty, Raj, Adam Looney, and Kory Kroft. 2009. "Salience and Taxation: Theory and Evidence." \textit{American Economic Review}, 99(4): 1145-77. \href{https://doi.org/10.1257/aer.99.4.1145}{https://doi.org/10.1257/aer.99.4.1145}.
\\ \\
Clean Vehicle Rebate Project. (2015, October 28). \textit{CVRP Participation Rates}. \href{https://cleanvehiclerebate.org/eng/content/cvrp-participation-thru-2015-03}{https://cleanvehiclerebate.org/eng/content/cvrp-participation-thru-2015-03}.
\\ \\
Clinton, B., Brown, A., Davidson, C., \& Steinberg, D. (2015). \textit{Impact of direct financial incentives in the emerging battery electric vehicle market: A preliminary analysis (presentation); nrel (national renewable energy laboratory)} (No. NREL/PR-6A20-63263). National Renewable Energy Lab.(NREL), Golden, CO (United States). \href{https://www.nrel.gov/docs/fy15osti/63263.pdf}{https://www.nrel.gov/docs/fy15osti/63263.pdf}.
\\ \\
Colias, M. \& Foldy, B. (2021). GM's Chevy Bolt Recall Casts Shadow Over EV Push. Retrieved 9 September 2021, from \href{https://www.wsj.com/articles/gms-chevy-bolt-recall-casts-shadow-over-ev-push-11630246736?st=eju6p7vejqsxiu4}{https://www.wsj.com/articles/gms-chevy-bolt-recall-casts-shadow-over-ev-push-11630246736?st=eju6p7vejqsxiu4}.
\\ \\
\textit{Compustat Industrial [Annual Data]}. (2010-2020). Available: Standard \& Poor's/Compustat [2021, August 15]. Retrieved from Wharton Research Data Service.
\\ \\
Covert, T., Greenstone, M., \& Knittel, C. R. (2016). Will we ever stop using fossil fuels?. \textit{Journal of Economic Perspectives, 30}(1), 117-38. \href{https://doi.org/10.1257/jep.30.1.117}{https://doi.org/10.1257/jep.30.1.117}.
\\ \\
Daziano, R. A., \& Bolduc, D. (2013). Incorporating pro-environmental preferences towards green automobile technologies through a Bayesian hybrid choice model. \textit{Transportmetrica A: Transport Science, 9}(1), 74-106. \href{https://doi.org/10.1080/18128602.2010.524173}{https://doi.org/10.1080/18128602.2010.524173}.
\\ \\
DeShazo, J. R. (2016). Improving incentives for clean vehicle purchases in the United States: challenges and opportunities. Review of Environmental Economics and Policy, 10(1), 149-165. \href{https://doi.org/10.1093/reep/rev022}{https://doi.org/10.1093/reep/rev022}.
\\ \\
DeVries, C. O. C. (2018, December 25). \textit{Electric car credits benefit the elite over the many}. TheHill. \href{https://thehill.com/opinion/finance/422732-electric-car-credits-benefit-the-elite-over-the-many}{https://thehill.com/opinion/finance/422732-electric-car-credits-benefit-the-elite-over-the-many}.
\\ \\
Dyni, J. R. (2006). Geology and resources of some world oil-shale deposits. \href{https://pubs.usgs.gov/sir/2005/5294/pdf/sir5294_508.pdf?simple=True}{https://pubs.usgs.gov/sir/2005/5294/pdf/sir5294\_508.pdf?simple=True}.
\\ \\
Energy Information Administration. (2021, March 18). \textit{Electricity explained}. \href{https://www.eia.gov/energyexplained/electricity/electricity-in-the-us.php}{https://www.eia.gov/energyexplained/electricity/electricity-in-the-us.php}.
\\ \\
Energy Information Administration. (2011, September 19). \textit{China and India account for half of global energy growth through 2035}. \href{https://www.eia.gov/todayinenergy/detail.php?id=3130}{https://www.eia.gov/todayinenergy/detail.php?id=3130}.
\\ \\
Erb, T. (2021, March 30). The Social Cost of Carbon - Going Nowhere But Up. Center for Climate and Energy Solutions. \href{https://www.c2es.org/2021/03/the-social-cost-of-carbon-going-nowhere-but-up/}{https://www.c2es.org/2021/03/the-social-cost-of-carbon-going-nowhere-but-up/}.
\\ \\
Federal EV Tax Credit Phase Out Tracker By Automaker - EVAdoption. (2021). Retrieved 23 September 2021, from \href{https://evadoption.com/ev-sales/federal-ev-tax-credit-phase-out-tracker-by-automaker/}{https://evadoption.com/ev-sales/federal-ev-tax-credit-phase-out-tracker-by-automaker/}.
\\ \\
Ford, C., 2021. \textit{Interpreting Log Transformations in a Linear Model | University of Virginia Library Research Data Services + Sciences}. [online] Data.library.virginia.edu. Available at: <\href{https://data.library.virginia.edu/interpreting-
log-transformations-in-a-linear-model/}{https://data.library.virginia.edu/interpreting-
log-transformations-in-a-linear-model/}> [Accessed 13 September 2021].
\\ \\
Gayer, T., \& Parker, E. (2013). Cash for Clunkers: An evaluation of the car allowance rebate system. \textit{Brookings Institution}, 9. \href{https://www.brookings.edu/wp-content/uploads/2016/06/cash_for_clunkers_evaluation_paper_gayer.pdf}{https://www.brookings.edu/wp-content/uploads/2016/06/cash\_for\_clunkers\_evaluation\_paper\_gayer.pdf}.
\\ \\
Gillingham, K. T., Houde, S., \& van Benthem, A. A. (2021). Consumer myopia in vehicle purchases: evidence from a natural experiment. American Economic Journal: Economic Policy, 13(3), 207-38. \href{https://doi.org/10.1257/pol.20200322}{https://doi.org/10.1257/pol.20200322}.
\\ \\
Goldin, J. (2015). Optimal tax salience. \textit{Journal of Public Economics, 131}, 115-123. \href{http://dx.doi.org/10.1016/j.jpubeco.2015.09.005}{http://dx.doi.org/10.1016/j.jpubeco.2015.09.005}.
\\ \\
H.R.2454 - 111th Congress (2009-2010): American Clean Energy and Security Act of 2009. (2009, July 7). \href{https://www.congress.gov/bill/111th-congress/house-bill/2454}{https://www.congress.gov/bill/111th-congress/house-bill/2454}.
\\ \\
H.R.3684 - 117th Congress (2021-2022): Infrastructure Investment and Jobs Act. (2021, October 1). \href{https://www.congress.gov/bill/117th-congress/house-bill/3684}{https://www.congress.gov/bill/117th-congress/house-bill/3684}.
\\ \\
Hardman, S., Chandan, A., Tal, G., \& Turrentine, T. (2017). The effectiveness of financial purchase incentives for battery electric vehicles-A review of the evidence. \textit{Renewable and Sustainable Energy Reviews, 80}, 1100-111. \href{http://dx.doi.org/10.1016/j.rser.2017.05.255}{http://dx.doi.org/10.1016/j.rser.2017.05.255}.
\\ \\
Heffner, R. R., Kurani, K. S., \& Turrentine, T. S. (2007). Symbolism in California's early market for hybrid electric vehicles. \textit{Transportation Research Part D: Transport and Environment, 12}(6), 396-413. \href{https://doi.org/10.1016/j.trd.2007.04.003}{https://doi.org/10.1016/j.trd.2007.04.003}.
\\ \\
Holland, S. P., Mansur, E. T., Muller, N. Z., \& Yates, A. J. (2020). Decompositions and policy consequences of an extraordinary decline in air pollution from electricity generation. \textit{American Economic Journal: Economic Policy, 12}(4), 244-74. \href{https://doi.org/10.1257/pol.20190390}{https://doi.org/10.1257/pol.20190390}.
\\ \\
Holland, Stephen P.; Mansur, Erin T.; Muller, Nicholas Z.; and Yates, Andrew J., "Are There Environmental Benefits from Driving Electric Vehicles? The Importance of Local Factors" (2016). Open Dartmouth: Faculty Open Access Articles. 2382. \href{https://doi.org/10.1257/aer.20150897}{https://doi.org/10.1257/aer.20150897}. 
\\ \\
Interagency Working Group. (2021). \textit{Technical Support Document: Social Cost of Carbon, Methane, and Nitrous Oxide Interim Estimates under Executive Order 13990.} Tech. rep., White House. URL \href{https://www.whitehouse.gov/wp-content/uploads/2021/02/TechnicalSupportDocument\_SocialCostofCarbonMethaneNitrousOxide.pdf}{https://www.whitehouse.gov/wp-content/uploads/2021/02/TechnicalSupportDocument\_SocialCostofCarbonMethaneNitrousOxide.pdf}.
\\ \\
Kelley Blue Book. (2021, September 14). \textit{New-Vehicle Prices Surge to Record Highs for Fifth Straight Month, According to Kelley Blue Book.} Kelley Blue Book | MediaRoom. Retrieved October 31, 2021, from \href{https://mediaroom.kbb.com/2021-09-14-New-Vehicle-Prices-Surge-to-Record-Highs-for-Fifth-Straight-Month,-According-to-Kelley-Blue-Book}{https://mediaroom.kbb.com/2021-09-14-New-Vehicle-Prices-Surge-to-Record-Highs-for-Fifth-Straight-Month,-According-to-Kelley-Blue-Book}.
\\ \\
Lambert, F. (2018). GM kills the Chevy Volt, shuts down factories, but accelerates EV investment. Retrieved 5 September 2021, from \href{https://electrek.co/2018/11/26/gm-chevy-volt-factory-shutdown-electric-investmet/}{https://electrek.co/2018/11/26/gm-chevy-volt-factory-shutdown-electric-investmet/}
\\ \\
Lambert, F. (2021). Tesla Model S caught fire and burned down while charging at a Supercharger [Gallery]. Retrieved 9 September 2021, from \href{https://electrek.co/2016/01/01/tesla-model-s-caught-fire-and-burned-down-charging-supercharger/}{https://electrek.co/2016/01/01/tesla-model-s-caught-fire-and-burned-down-charging-supercharger/}.
\\ \\
Layard, R., Mayraz, G., \& Nickell, S. (2008). The marginal utility of income. \textit{Journal of Public Economics, 92}(8-9), 1846-1857. \href{https://doi.org/10.1016/j.jpubeco.2008.01.007}{https://doi.org/10.1016/j.jpubeco.2008.01.007}.
\\ \\
Leard, B. L. (2021, March 24). \textit{Federal Climate Policy 104: The Transportation Sector} [Illustration]. \href{https://www.rff.org/publications/explainers/federal-climate-policy-104-the-transportation-sector/}{https://www.rff.org/publications/explainers/federal-climate-policy-104-the-transportation-sector/}.
\\ \\
Levitt, S., List, J., \& Syverson, C. (2013). Toward an Understanding of Learning by Doing: Evidence from an Automobile Assembly Plant. \textit{Journal of Political Economy, 121}(4), 643-681. \href{https://doi.org/10.1086/671137}{https://doi.org/10.1086/671137}.
\\ \\
Li, S., Linn, J., \& Spiller, E. (2013). Evaluating "Cash-for-Clunkers": Program effects on auto sales and the environment. \textit{Journal of Environmental Economics and management, 65}(2), 175-193. \href{https://doi.org/10.1016/j.jeem.2012.07.004}{https://doi.org/10.1016/j.jeem.2012.07.004}.
\\ \\
Li, S., Tong, L., Xing, J., \& Zhou, Y. (2017). The market for electric vehicles: indirect network effects and policy design. Journal of the Association of Environmental and Resource Economists, 4(1), 89-133. \href{http://dx.doi.org/10.1086/689702}{http://dx.doi.org/10.1086/689702}.
\\ \\
Modaresi, R., Pauliuk, S., Løvik, A. N., \& Müller, D. B. (2014). Global carbon benefits of material substitution in passenger cars until 2050 and the impact on the steel and aluminum industries. \textit{Environmental science \& technology, 48}(18), 10776-10784. \href{https://doi.org/10.1021/es502930w}{https://doi.org/10.1021/es502930w}.
\\ \\
Ritchie, H., \& Roser, M. (2020) - "$\mathrm{CO}_{2}$ and Greenhouse Gas Emissions". \textit{Published online at OurWorldInData.org.} Retrieved from: '\href{https://ourworldindata.org/co2-and-other-greenhouse-gas-emissions}{https://ourworldindata.org/co2-and-other-greenhouse-gas-emissions}' [Online Resource]
\\ \\
Roster, E. (2021, January 22). The Most Important Number You've Never Heard Of. Bloomberg Green. \href{https://www.bloomberg.com/news/articles/2021-01-22/how-do-you-put-a-price-on-climate-change-michael-greenstone-knows?sref=pD2ECzY4}{https://www.bloomberg.com/news/articles/2021-01-22/how-do-you-put-a-price-on-climate-change-michael-greenstone-knows?sref=pD2ECzY4}.
\\ \\
S.395 - 117th Congress (2021-2022): Electric CARS Act of 2021. (2021, February 23). \href{https://www.congress.gov/bill/117th-congress/senate-bill/395}{https://www.congress.gov/bill/117th-congress/senate-bill/395}.
\\ \\
Senator Deborah Fischer. (2021, August 11). \textit{Fischer Opposes Senate Democrats' Reckless Tax and Spend Plan} [Press release]. \href{https://www.fischer.senate.gov/public/index.cfm/news?ID=C9544E5D-4703-4DFD-97B1EB36A7D64814}{https://www.fischer.senate.gov/public/index.cfm/news?ID=C9544E5D-4703-4DFD-97B1EB36A7D64814}.
\\ \\
Text - S.2118 - 117th Congress (2021-2022): Clean Energy for America Act. (2021, June 21). \href{https://www.congress.gov/bill/117th-congress/senate-bill/2118/text}{https://www.congress.gov/bill/117th-congress/senate-bill/2118/text}.
\\ \\
The White House (2021, August 5) FACT SHEET: President Biden Announces Steps to Drive American Leadership Forward on Clean Cars and Trucks. \href{https://www.whitehouse.gov/briefing-room/statementsreleases/2021/08/05/fact-sheet-president-biden-announces-steps-to-drive-american-leadership-forward-on-cleancars-and-trucks/}{https://www.whitehouse.gov/briefing-room/statementsreleases/2021/08/05/fact-sheet-president-biden-announces-steps-to-drive-american-leadership-forward-on-cleancars-and-trucks/}.
\\ \\
The White House (2021, July 28) FACT SHEET: Historic Bipartisan Infrastructure Deal. \href{https://www.whitehouse.gov/briefing-room/statements-releases/2021/07/28/fact-sheet-historic-bipartisan-infrastructure-deal/}{https://www.whitehouse.gov/briefing-room/statements-releases/2021/07/28/fact-sheet-historic-bipartisan-infrastructure-deal/}.
\\ \\
United States Environmental Protection Agency. (2021). Sources of Greenhouse Gas Emissions. \href{https://www.epa.gov/ghgemissions/sources-greenhouse-gas-emissions}{https://www.epa.gov/ghgemissions/sources-greenhouse-gas-emissions}.
\\ \\
Wang, S \& Ge, M. (2019, October 16). Everything You Need to Know About the Fastest-Growing Source of Global Emissions: Transport. World Resources Institute. \href{https://www.wri.org/insights/everything-you-need-know-about-fastest-growing-source-global-emissions-transport}{https://www.wri.org/insights/everything-you-need-know-about-fastest-growing-source-global-emissions-transport}.
\\ \\
Ziegler, A. (2012). Individual characteristics and stated preferences for alternative energy sources and propulsion technologies in vehicles: A discrete choice analysis for Germany. \textit{Transportation Research Part A: Policy and Practice, 46}(8), 1372-1385. \href{https://doi.org/10.1016/j.tra.2012.05.016}{https://doi.org/10.1016/j.tra.2012.05.016}.
\\ \\
Ziegler, M. S., \& Trancik, J. E. (2021). Re-examining rates of lithium-ion battery technology improvement and cost decline. \textit{Energy \& Environmental Science, 14}(4), 1635-1651. \href{https://doi.org/10.1039/D0EE02681F}{https://doi.org/10.1039/D0EE02681F}.
\end{sloppypar}

\section{Appendix}
\subsection{A. Role of Electric Vehicles in Promoting Human Welfare}

Sometimes regarded as "the most important number you've never heard of," the social cost of carbon (SCC) is an important tool used by policymakers to compare the cost and benefits of environmental policies (Roster, 2021)

The average net social benefit per EV can be estimated by multiplying the difference in lifetime carbon emissions between the average EV and ICE by the SCC (Aguirre et al., 2012).\footnote{Using a life cycle assessment (LCA), Aguirre et al. (2012) find the difference in lifetime carbon emissions between an EV and an ICE to be $31.045$ metric tons.} The low bound net social benefit is $\$ 1,583$ (using the White House's $3 \%$ discount rate $\mathrm{SCC}$ of $\$ 51$ per metric ton), while the high bound net social benefit is $\$ 3,881$ (using Carleton and Greenstone's $2 \%$ discount rate SCC of $\$ 125$ per metric ton) (Interagency Working Group, 2021; Carleton and Greenstone, 2021). Applying the average treatment effect obtained in this study, the social opportunity cost per EV model in 2019 is $\$ 14.6$ million in the low SCC scenario, and \$35.9 million in the high SCC scenario.\footnote{The average effect found in the initial estimate, 9,254, is determined by averaging the effect of the federal subsidy on sales of the Tesla Model S and Chevy Volt. It does not necessarily represent the average effect the subsidy would have on sales of every model in the EV market in 2019.} The global net social impact is dramatically higher. According to UN estimates, some $1.2$ billion new cars will be added to the global fleet by 2050. Assuming that all new cars are ICEs and that the net social benefit per EV remains constant at the low bound $(\$ 1,583)$, the global social opportunity cost is around $\$ 1.9$ trillion (Modaresi, 2014).

The estimates of the net social benefit per EV suggest a reduction of the maximum credit $(\$ 7,500)$ by $\$ 5,917$ (using the low estimate) and $\$ 3,619$ (using the high estimate). Though, at least three contingent factors indicate that a reduction may not be optimal. First, an LCA narrowly measures environmental impact in terms of carbon emissions, excluding other relevant factors such as nitrogen oxide and volatile organic compound emissions. An LCA that accounts for a broader array of environmental factors could indicate a greater EV net environmental benefit than current estimates suggest. Second, current SCC estimates exclude an increasingly important factorequity. The same nominal damage from climate change clearly has different impacts on countries of varying prosperity.\footnote{Hurricanes Dorian and Harvey illustrate how a climate disaster can have markedly different relative impacts on regions of varying wealth. Dorian, which made landfall in the relatively poor island nation of the Bahamas, costed \$3.4 billion (a significant but not exceptional figure). Despite the moderate nominal damage, the effect of the hurricane on the islands of Grand Bahama and Great Abaco in particular are strikingly visible today. Meanwhile, Harvey caused around 37 times more damage than Dorian (\$125 billion), but evidence of its existence has all but disappeared since it made landfall in Texas (a relatively wealthy U.S state). See Blake and Zelinsky, 2018; Damages and other impacts on Bahamas by Hurricane Dorian Estimated at $\$ 3.4$ billion: report, 2019 .} If wealth and income inequality continue to grow between the so-called global north and south, which I expect it will, then the impact of climate change will become even more disparate on a country level comparison (not to mention the growing income inequality within many developed and developing countries). An equity-weighted approach would account for disproportionate climate implications between regions. If implemented, it could raise the SCC by a factor of $2.5$ or more (Anthoff and Johannes, 2016). Third, the SCC does not account for the cost of consumer responses to growing climate risk. The increasing frequency and intensity of extreme weather events would likely lead consumers to purchase preventative measures such as insurance (Erb, 2021). These factors reduce confidence in the current EV net social benefit. Given the uncertainty of the SCC and EV net environmental benefit, it may not be optimal to lower the maximum credit of the federal subsidy until more comprehensive estimates are available.

\subsection{B. Alternative Policy Design}
Although this study finds that the current federal income tax subsidy substantially increases EV sales, alternative policy designs might be more cost-effective.

\subsubsection{I. EV Charging Infrastructure}
The slow charging time of EVs on standard household plugs has greatly stimulated demand for public fast-charging stations. Despite the integral role of access to fast charging in consumer purchase decisions, these stations have been underproduced because of a free-rider problem: individual firms incur significant costs to set up and maintain extensive public charging infrastructure, but benefits (i.e., access to charging) can be easily exploited by consumers of other firms that do not pay (i.e., free riders). If automakers act rationally, there will be a lack of charging stations since firms do not want to shoulder the private costs of building charging stations while being forced to share in the benefits.\footnote{Tesla has evaded this market failure by creating a global network of proprietary "supercharger" stations. See \href{https://www.tesla.com/supercharger}{https://www.tesla.com/supercharger}. This network makes charging more convenient for Tesla customers but may waste resources as it duplicates the number of necessary charging stations.} Despite the attempt by third party firms to bridge this supply gap, the charging fees they impose on consumers raises the cost of owning an EV and diminishes the expected value of fuel savings - the primary cost advantage of owning an EV over owning an ICE. Federal intervention could attempt to solve this free rider problem through direct construction or subsidization of a nationwide network of free public fast charging stations.

Li et al. (2017) estimate that subsidizing charging infrastructure could be twice as effective at encouraging EV consumption than the current federal income tax subsidy. This provides strong motivation for the federal government to subsidize construction of EV charging infrastructure. The Infrastructure Investment and Jobs Act does just that, allocating over $\$ 7$ billion in federal funds to EV infrastructure. Even so, the true effect of such investment on EV sales remains unclear.

\subsubsection{II. Income Tiered Incentives}
An income stratified subsidy that favors lower income consumers is more politically palatable than the federal subsidy. It also increases marginal environmental benefits. This is because, holding its environmental footprint constant, an EV replacing a less fuel-efficient ICE creates a greater marginal environmental benefit than if it were replacing a more fuel-efficient ICE. Bhat, Sen, and Eluru (2009) find that lower-income consumers tend to drive vehicles that are older, more polluting, run on fossil fuels, and drive them longer than higher-income consumers. An income stratified version of the federal subsidy reflects this higher likelihood. For example, low-income consumers could receive a maximum credit greater than the current maximum of $\$ 7,500$, and high income consumers receive a maximum credit less than $\$ 7,500$. While a recent amendment to the $\$ 3.5$ trillion Democratic infrastructure proposal put forth by Senator Deborah Fischer (R-Neb.) aims to address the issue of equity in the federal subsidy,\footnote{Senator Deborah Fischer's non-binding amendment to the current $\$ 7,500$ federal tax credit would eliminate the federal subsidy for all EVs priced over $\$ 40,000$ and for households with an income of $\$ 100,000$ or more. See press release: Fischer Opposes Senate Democrats' Reckless Tax and Spend Plan, 2021.} it differs fundamentally from the framework suggested here.\footnote{Notably, the Bloomberg NEF Electric Vehicle Outlook predicts that the average price of most EVs will fall below $\$ 40,000$ around 2030, implying that Senator Fischer's amendment may be a decade too early.}

\subsubsection{III. Salience in Consumer Purchase Decisions}
A central assumption in public finance is that agents fully optimize their behavior with respect to taxes. The current federal income tax credit performs optimally under this assumption. However, recent literature on salience demonstrates that agents are inattentive or myopic to some forms of taxes (Goldin, 2015; Chetty, Looney, and Kroft, 2009; Finkelstein, 2009). In the context of the federal subsidy, this literature suggests that policymakers should prioritize more salient subsidies such as sales tax reductions and rebates over the existing federal income tax credit.

Another factor in consumer purchase decisions related to salience is redemption cost, which varies based on the complexity of the subsidy and the labor required to obtain it. Sales tax reductions and fee exemptions require no redemption costs as they can be made available at the point of sale. Meanwhile, the relatively high redemption cost of less salient incentives such as rebates and tax credits diminish their expected value for consumers. The combined effect of high redemption costs and low salience may be illustrated by the relatively small uptake rate of California's state EV rebate program.\footnote{As of 2015, there was a $74 \%$ redemption rate for California's state EV rebates (currently valued up to $\$ 7,000$) (see Clean Vehicle Rebate Project, 2015). I use California's rebate in lieu of the federal income tax credit since there is scarce evidence for the uptake rate of the federal subsidy. This is appropriate because $42 \%$ percent of U.S. EVs are purchased in California, and rebates and tax credits are similar in terms of salience (see Electric Vehicle Registrations by State, 2020).} However, there is scarce empirical evidence for the relationship between low uptake rates and EV sales, and if the low uptake rates are due to low salience or high redemption costs.

\subsection{C. Descriptive Statistics}
\begin{center}
\textbf{Table 5.} Regression Model Descriptive Statistics.
\vspace{1em}

\begin{tabular*}{\columnwidth}{@{\extracolsep{\stretch{1}}}*{8}{r}@{}}
\toprule[1.5pt] \\
Independent Variable & Min & Q1 & Median & Mean & Q3 & Max & SD \\ \\
\hline \\
Nominal Sales & 0 & 0 & 9 & 3,368 & 2,015 & 154,840 & 10,729.30 \\ \\
Logarithmic Sales & 0 & 0 & 2.303 & 3.803 & 7.609 & 11.950 & 3.89 \\ \\
\bottomrule[1.5pt] \\
\end{tabular*} \\
\end{center}

\noindent
Note: In the nominal sales data, the mean $(3,368)$ is about $374.22$ times larger than the median (9), indicating a significant right skew. In the logarithmic sales data, the mean (3.803) is about $1.65$ times greater than the median (2.303). The mean to median ratio of the logarithmic sales data is considerably smaller than that of the nominal sales data, indicating the logarithmic sales data has a more normal distribution. The primary data points influencing the nominal sales distribution's right skew are sales of the Tesla Model 3 in 2018 and $2019 (139,782$ and 154,840, respectively). Its sales in $2018 (139,782)$ is about $12.6$ standard deviations above the mean, and its sales in 2019 $(154,840)$ is about $14.4$ standard deviations above the mean.

\end{document}